\documentclass[twocolumn]{aastex631}
\begin{document}

\title{ASR: Archival Solar flaRes catalogue}

\shortauthors{Berretti \& Mestici et al.}
\author[0009-0007-2465-1931]{M. Berretti}
\correspondingauthor{M. Berretti}
\affiliation{University of Rome Tor Vergata, Department of Physics, Via della Ricerca Scientifica 3, 00133 Rome, Italy}
\affiliation{Univeristà di Trento, Via Calepina 14, 38122 Trento, Italy}
\email{michele.berretti@unitn.it}

\author[0009-0000-2442-8878]{S. Mestici}
\affiliation{University of Rome "La Sapienza", Department of Physics, P.le A. Moro 5, 00185 Rome, Italy}

\author[0000-0001-7369-8516]{L. Giovannelli}
\affiliation{University of Rome Tor Vergata, Department of Physics, Via della Ricerca Scientifica 3, 00133 Rome, Italy}

\author[0000-0003-2500-5054]{D. Del Moro}
\affiliation{University of Rome Tor Vergata, Department of Physics, Via della Ricerca Scientifica 3, 00133 Rome, Italy}

\author[0000-0002-5365-7546]{M. Stangalini}
\affiliation{ASI Italian Space Agency, Via del Politecnico snc, 00133 Rome, Italy}

\author[0000-0002-9691-8910]{F. Giannattasio }
\affiliation{Istituto Nazionale di Geofisica e Vulcanologia, Via di Vigna Murata 605, 00143 Roma, Italy}

\author[0000-0002-2276-3733]{F. Berrilli}
\affiliation{University of Rome Tor Vergata, Department of Physics, Via della Ricerca Scientifica 3, 00133 Rome, Italy}\email{francesco.berrilli@roma2.infn.it}

\begin{abstract}

Solar flares result from the rapid conversion of stored magnetic energy within the Sun's corona. These energy releases are associated with coronal magnetic loops, which are rooted in dense photospheric plasma and are passively transported by surface advection. Their emissions cover a wide range of wavelengths, with soft X-rays being the primary diagnostic for the past fifty years. Despite the efforts of multiple authors, we are still far from a complete theory, capable of explaining the observed statistical and individual properties of flares. Here, we exploit the availability of stable and long-term soft x-ray measurements from NASA's GOES mission to build a new solar flare catalogue, with a novel approach to linking sympathetic events. Furthermore, for the most energetic events since 2010, we have also provided a method to identify the origin of the observed flare and eventual link to the photospheric active region by exploiting the array of instruments onboard NASA's Solar Dynamic Observatory. Our catalogue provides a robust resource for studying space weather events and training machine learning models to develop a reliable early warning system for the onset of eruptive events in the solar atmosphere.
\end{abstract}


\section{Introduction} \label{sec:intro}

It is well known that space weather events such as solar flares, coronal mass ejections (CMEs), and solar energetic particle (SEP) emissions, pose a significant threat to human infrastructures both on Earth and in space \citep[e.g.][]{macalester_extreme_2014, Berrilli2014, DiFino2014, oughton_quantifying_2017}. Hence, understanding and predicting the hazards linked to the Sun's magnetic activity have grown in importance over the years. In the last decade, a plethora of authors have trained different machine learning models to forecast the likelihood that the Sun will produce a flare in a precise time frame, usually ranging from a few hours to 1-2 days \citep[][to name a few]{Schrijver2007, 2007SoPh..241..195Q, 2010ApJ...709..321Y, 2011ApJ...735..130L, 2015ApJ...798..135B, Barnes2016, 2016AGUFMSH34A..02J, 2018ApJ...856....7H, 2018ApJ...869...91P, 2019ApJ...881..101L, 2021EP&S...73...64N, 2021ApJ...922..232D, Cicogna2021,2023mlkd.conf...72P}. For an extensive analysis of the challenges of machine learning applied to space weather and the prediction of the occurrence of solar flares, we refer the reader to the review work of \cite{camporeale_challenge_2019}.

\begin{figure*}[ht!]
    \centering
    \includegraphics[width=\linewidth]{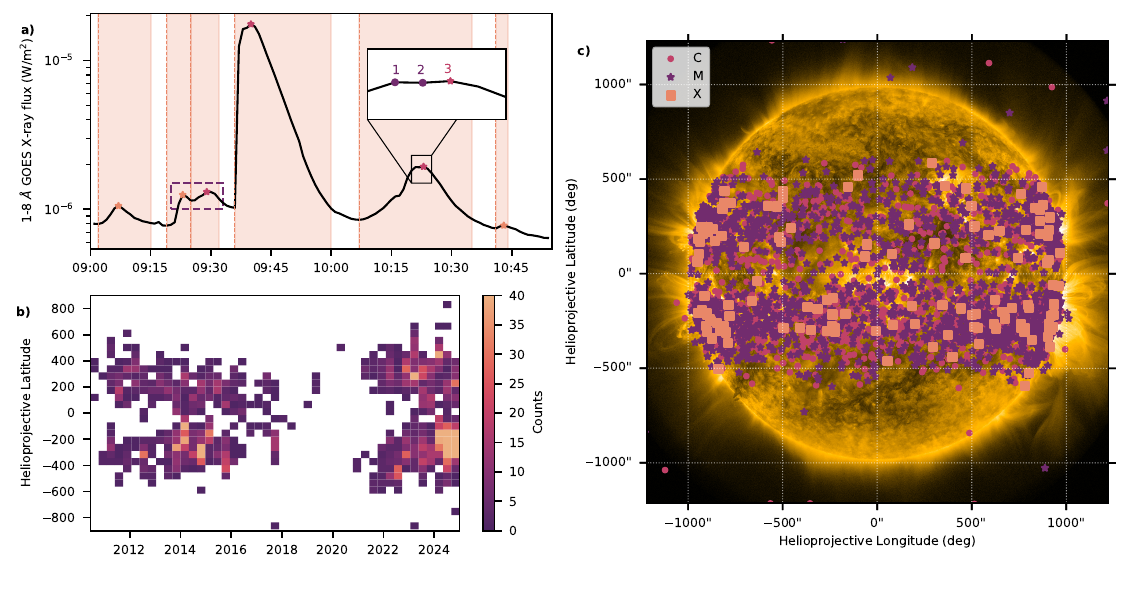}
    \caption{(a) Application of the flare detection algorithm to GOES X-ray data from 9 March 2015. The violet dashed rectangle highlights two homologous events correctly detected by the algorithm, while the magnified inset shows the difficulty of flare detection due to noise-induced fluctuations. Orange stars mark the flare peaks detected by our algorithm, while pink ones highlight those identified in both the NOAA catalogue and our own. Orange-shaded areas denote the duration of the corresponding flare. (b) Reconstructed butterfly diagram using solar flares, showing the latitudinal spread of flare occurrences over different solar cycles. (c) Spatial distribution of flares on the solar disk, with flare types represented by colour. Stray values are due to oversaturated AIA images.}
    \label{fig:1}    
\end{figure*}

A complete data set is crucial for enhancing the performance of machine learning models while also mitigating biases and optimising computational efficiency, as discussed in \cite{mehrabi_bias_2021}. Traditionally, many works on flare forecasting are based on the flare catalogue provided by the NOAA\footnote{\url{ftp://ftp.swpc.noaa.gov/pub/warehouse/}}, a live service that has provided flare events since January 2002 with the launch of the GOES-08 satellite. However, while the NOAA catalogue remains the primary reference for flaring events, standard flare identification methods can miss some events and fail to provide further insights into the more complex processes linking flares to the global-scale dynamics of the solar magnetic field. Alternative catalogues have been proposed, often relying on different detection algorithms or instruments.\\
In \cite{Aschwanden2012} the authors conduct a comprehensive analysis of solar flare activity spanning 37 years (1975-2011). Using an automated flare detection algorithm applied to soft X-ray data from GOES satellites, they identified and catalogued over 300,000 solar flare events. From now on, we will refer to this catalogue as ASC.\\
In \cite{2022FrASS...931211V}, the authors present a solar flare catalogue based on EUV images from the Atmospheric Imaging Assembly \citep[AIA,][]{2012SoPh..275...17L} onboard the Solar Dynamics Observatory \citep[SDO,][]{2012SoPh..275....3P}. This approach increases the number of recorded events and improves the characterisation of the total energy released. However, its applicability is limited by the relatively short operational window of the SDO mission compared to GOES, restricting statistical studies across multiple solar cycles.\\
In \cite{plutino2023new}, which has been referred to as the PLU catalogue from here on, the authors propose an alternative flare identification algorithm to detect events in the GOES X-ray signal. Instead of relying on the ratio between the background and the peak fluxes, their approach defines a true event based on an instrumental threshold, leading to a higher number of detected flares.

In this work, we present a new solar flare catalogue based on soft X-ray measurements from GOES and named Archival Solar flaRes (ASR) catalogue. We refine many of the events already recorded by both NOAA and PLU and employ a novel approach to identify and connect both sympathetic flares and homologous flares. This analysis allows us to investigate statistical relationships between distant flaring regions and explore the global properties of these energetic solar events.
Thanks to the availability of stable and long-term X-ray measurements, we are able to cover two full solar cycles, and, for the most recent one, we also provide the origin point of the most intense events. Our catalogue serves as a valuable resource for both space weather studies and machine learning applications. Indeed, thanks to a versatile and complete data set, we can support the development of more reliable flare prediction models and early warning systems for potentially disruptive solar activity.

\section{Dataset}

The Geostationary Operational Environmental Satellites (GOES), first launched in 1974 with GOES-1 and currently active with GOES-16/17/18, is a NASA/NOAA mission focused on monitoring both Earth and solar activity.
Regarding the Sun observation side of the mission, the satellites are equipped with two X-ray sensors (XRS), one measuring radiation between 0.5 and 4 \textup{~\AA} (short channel) and the other between 1 and 8 \textup{~\AA} (long channel). See \cite{Hanser1996} for an introduction to GOES XRS.
In this work, we considered the 1-minute averaged science-quality X-ray signal of the long channel of the primary satellite from January 1, 2002, to December 31, 2024.
The science quality data, provided directly by the NOAA science team, is thoroughly processed and includes several corrections with respect to the operational data, including those for SWPC (Space Weather Prediction Centre) scaling factors,  XRS-A bandpass, time stamps, temperature effects, better quality flags, and other issues. Specific information on the available data for each satellite and which to consider for a specific time range can be found in the extensive guide provided by the NOAA in the dedicated XRS Web portal\footnote{\url{https://www.ngdc.noaa.gov/stp/satellite/goes-r.html}}. In general, when two or more satellites are actively collecting data at the same time, one of them is indicated as the primary source, while the other is indicated as the secondary one. Finally, the downloader function of the ASR catalogue code provides yearly datasets containing the 1-minute averaged time series of the soft x-ray flux along with the quality flag, the data source, and the time stamp in plain text format. With this approach, our method differs from the previous ASC and PLU catalogues, which use the GOES operational data, characterised by a temporal resolution ranging between 1 and 4 seconds later averaged to 12 seconds and with an additional smoothing window applied.

For the most energetic events, we are able to determine the origin point of the solar eruption thanks to the availability of stable, seeing-free, and continuous images of the solar corona captured by the AIA onboard SDO. To determine the location of solar flares identified through our GOES X-ray analysis, we chose to utilise images from Fe~{\sc ix}~$17.1$~nm spectral line, corresponding to images of the lower quiet corona to provide a reliable picture of the coronal magnetic fields, with little interference from other transient events such as shocks (\cite{van2022solar}). Here, we estimated the position of 138 X-class flares, 1735 M-class flares and 1958 C-class flares, for a total of 3831 events.

\begin{figure}
    \centering
    \includegraphics[width=\linewidth]{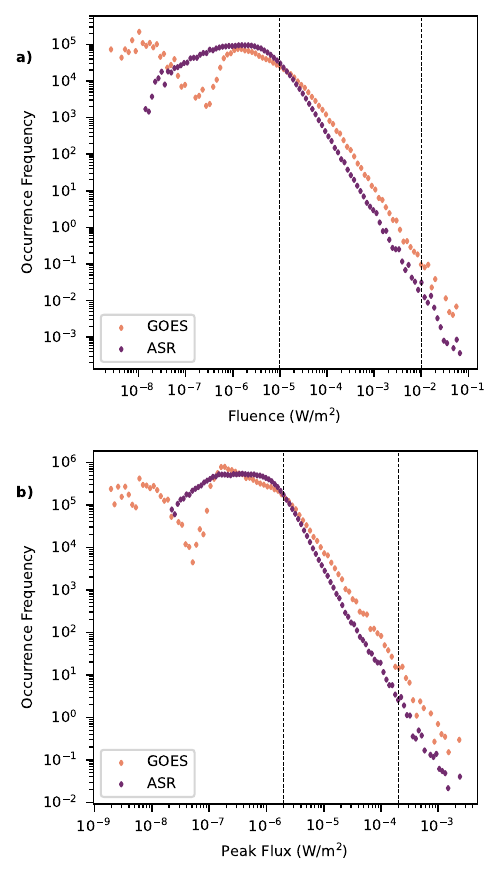}
    \caption{Probability density functions of the estimated flux integral, also known as fluence, (top) and peak flux (bottom) for the GOES catalogue and the ASR catalogue, in orange and violet, respectively. The dashed black lines highlight the range used for the linear fit to estimate the slope of the distributions. In the top panel, this range spans from $3\cdot10^{-5}$ to $10^{-2}$ W/m$^2$ , and the fitted slopes are $-2.1\pm0.1$ and $-1.9\pm0.01$ respectively for the ASR and GOES ones. In the bottom panel, it extends from $3\cdot10^{-6}$ to $2\cdot10^{-4}$ W/m$^2$, and the estimated slopes are $-2.3\pm0.01$ for the ASR catalogue and $-2.1\pm0.01$ for the GOES one.}
    \label{fig:2}
\end{figure}

\section{Methods}

The ASR catalogue code consists of a significant improvement over the previous PLU flare catalogue, with several of the same authors contributing to both works. Although many of the novel features of the former catalogue have been inherited, such as accounting for sympathetic/homologous events and computing each flare's flux integral, the ASR catalogue further offers many improvements. In this section, we will thoroughly review the various steps involved in the identification, definition, and characterisation of flares in our catalogue.

\subsection{Flare Identification}

The first step in building the flare catalogue consists of identifying maxima and minima in the yearly GOES time series. To this end, we exploited the ``argrelextrema" function of the SciPy library (\cite{2020SciPy-NMeth}). For each sampled point, the function checks its nearest neighbours; if both values are larger or smaller than the considered point, then it is classified as a relative minimum or maximum, respectively. The index of the detected extreme is then recorded in a ``pandas dataframe" data structure and saved as a separate CSV (Comma-Separated Values) file. A drawback of this approach is the loss of flat extremes, that is extremes characterised by a series of the same value. However, such situations are not commonly found in the soft X-ray signal and are mainly associated with instrumental saturation. Hence, saturated events are handled by replacing the flat region in the measured GOES X-ray signal with a second-degree polynomial fitted by overestimating the peak by $2.5\%$ in the centre of the saturated region and adjusting the edges. It is worth noting that, in the 22 years covered by the whole catalogue, instrumental saturation occurred only once, during the famous storm of October 2003. \\
Due to the inherently noisy nature of the GOES signal, even in the softer long channel and 1-minute averaged science-quality data, we need to account for the presence of consecutive extrema. These neighbouring extrema are rarely associated with true flare events, which are generally characterised by a sharp increase in flux followed by a slower energy decay. Thus, considering these extrema as events would lead to the presence of false flares, compromising the statistics and goodness of the catalogue. In our code, this is taken into account by checking the indexes of the extrema: if three consecutive ones are detected, the first two are dropped. We point out that this might lead to an offset in either the flare start, end, or peak time of no more than two minutes.

\subsection{Flare Definition}

Once the relative extrema of the entire GOES time series are identified, we check how many of these events are true flares and not noise-driven fluctuations. Over the years, various definitions have been proposed to reliably identify solar flares in the GOES signal. The NOAA catalogue defines a flare using three criteria: 1) the event starts where four consecutive points increase monotonically and exceed $10^{-7}$ W/m$^2$; 2) the flux ratio between the first (start) and fourth (peak) points is greater than 1.4; 3) the end time of the flare is defined as the point in time successive to the flare peak where the flux is half that observed at the peak.

In our catalogue, we adopt the same flare definition as in PLU. Each minimum represents a potential flare start, while the following maximum represents the possible flare peak. A provisional flare event is classified as a true flare if the flux at the peak exceeds the background flux, i.e. the averaged flux two minutes prior to the minimum, by a threshold equal to $2\cdot10^{-8}$ W/m$^2$. Properly estimating the flux background can greatly improve the detection of weaker events and the estimation of their energy. The threshold value, related to the intrinsic sensibility of the XRS instrument, defines the noise floor for typical GOES flux measurements, established during quiescent periods of solar activity \citep[e.g.,][]{Aschwanden2012}.
Accurately estimating the background flux is crucial. Indeed, this flux can be considered as the superposition of the contributions coming from the active region where the flare originates, as well as the integrated flux coming from the entire solar disk (including beyond the visible limb). \\
We define the end time of an event either as the time successive to the peak where the flux value returns to the background value plus the threshold value, or the point when a new true event starts. The choice between the two falls to whichever occurs first. In the former case, a unique label is assigned to the flare. In the latter case, that is, in case of consecutive true flare events, a shared label is assigned to the group of events, allowing direct identification of homologous or sympathetic events. 

\subsection{Flare Characterisation}

In addition to the start, peak, and end times and their associated flux values, our catalogue offers other properties related to the flare that can be used to characterise each event. The integrated flux of each flare is computed as the integral, using the trapezoidal rule, of the flux values between the start and end times. This information is critical to understand the energy released during the entire flare event. We also provide a corrected integrated flux value that corresponds to the difference between the integrated and background fluxes. This quantity takes into account the background solar activity and offers insight into the "rescaled" energy released. 

To further compare our catalogue with the one released by the NOAA, we also provide the classification of flares according to the NOAA standard procedure. Indeed, flares are traditionally divided into five classes depending on the intensity of the peak emission in the soft X-ray signal, starting from the weakest A-class flares, with peak flux emissions between $10^{-8}$ and $10^{-7}$ W/m$^2$ up to X-class flares with peak emissions over $10^{-4}$ W/m$^2$. Similarly to the flux integral, we provide two different class definitions for any flare event. The traditional class is determined similarly to the NOAA catalogue by considering the peak flux of the flare event. We then introduce a relative flare class that is estimated by rescaling the event peak flux to the background flux (referred to as delta flux from here on), so that it is not biased by the solar activity cycle. A detailed description of the estimated background is provided in Appendix \ref{sec:app}. For the sake of clarity, from now on we will refer to the catalogue which uses the traditional class definition as the ASR catalogue and to the one using the relative class definition as the ASR$_{REL}$ catalogue.

Finally, to align with the NOAA procedure and enable further refinement of observed flaring events, we also include the ratio between peak and background flux. Adjusting the filters in our catalogue makes it easy to reconstruct the NOAA one.

\subsection{Flare Location} \label{Location}

For the most energetic events, that is, those with a delta flux greater than $5\cdot10^{-6}$ W/m$^2$, we estimate the point of origin of the flare. To determine the position of a given event, we consider the barycentre of the brightest cluster of pixels in the difference AIA Fe~{\sc ix}~$17.1$~nm channel image between the peak and start of the flare.  Moreover, as we consider only the largest and brightest cluster of pixels, we can safely assume that other energetic events in the solar atmosphere, such as weaker flares and shocks, can be neglected and do not risk affecting the flare localisation process. The flare location is then provided in pixel, helioprojective, and heliographic coordinates to allow for a straightforward connection to the corresponding active region.

\section{Results and discussion}

\begin{table}
\caption{Summary of the flare statistics for the NOAA, Plutino and ASR catalogues divided by class up to the year 2022.}
\label{table:1}
\centering
\begin{tabular}{c c c c c}
\hline\hline
Class & GOES & PLU & ASR & ASR$_{REL}$\\
\hline
A & 2690 & 2800 & 5077 & 147696\\
B & 15980 & 88005 & 118384  & 93963\\
C4- & 8543 & 68980 & 125051 & 17248\\
C5+ & 1564 & 4807 & 10352 & 2698\\
M & 1581 & 2803 & 5189 & 2490\\
X & 159 & 187 & 274 & 232\\
\hline
\end{tabular}
\end{table}

The total number of flares detected by our algorithm as a function of the flare class is shown in Table 1. In particular, the first, second, and third columns refer to the number of flares identified by the NOAA, PLU, and ASR catalogues, respectively. Instead, the fourth column reports the number of flares per class after accounting for the background flux, based on a ``relative" flux value obtained by subtracting the background flux from the peak one. We wish to stress that the background we intended here is related to the magnetic activity of the Sun and is not related to the background given by other galactic sources, which is already removed in the science-quality data provided by NOAA. The number of flares in our catalogue is consistently higher with respect to the NOAA one in each class. These differences, especially in the lower energy classes, stem from the radically different detection method used in our work, which is based on an instrumental threshold and does not require a specific ratio between the flux measured at the start and peak points. The increase in flares in the M and X classes can also be attributed to our definition of a flare. In fact, secondary peaks in the tail of M- or X-class events may also be classified with a similar class. Table 1 also shows that the number of flares identified in our catalogue is also greater than that presented in PLU. Although the identification and definition processes are the same, the main difference lies in the dataset used. In fact, the former uses the operational data provided by GOES, which requires further processing. 
We also observe that the relative class attribution, which takes into account the background solar activity, drastically changes the distribution of flares, especially at lower energies. In fact, most of the standard flares populating the B and C classes are shifted to a lower-energy class in the adjusted flux integral scenario. The X and M classes are instead less affected by the flux rescaling, meaning that most of the large events, which are generally observed during high solar activity, largely overshoot the background-level flux. 

It is worth pointing out that the number of flare events for a catalogue is not a significant proxy of its quality. Indeed, this number is mainly related to the flare event definition, which can be quite arbitrary. Here, we aimed to provide the most general and versatile definition of a flare, leaving the users the possibility and freedom to decide which events to consider and which not. Furthermore, the grouping of homologous and/or sympathetic events under the same id offers the opportunity to consider these events as a single longer event or multiple separated ones. We wish to stress yet again that the novelty and strength of this work lies in its versatility, making it easily adaptable to many different tasks.

\begin{figure*}[ht!]
    \centering
    \includegraphics[width=\linewidth]{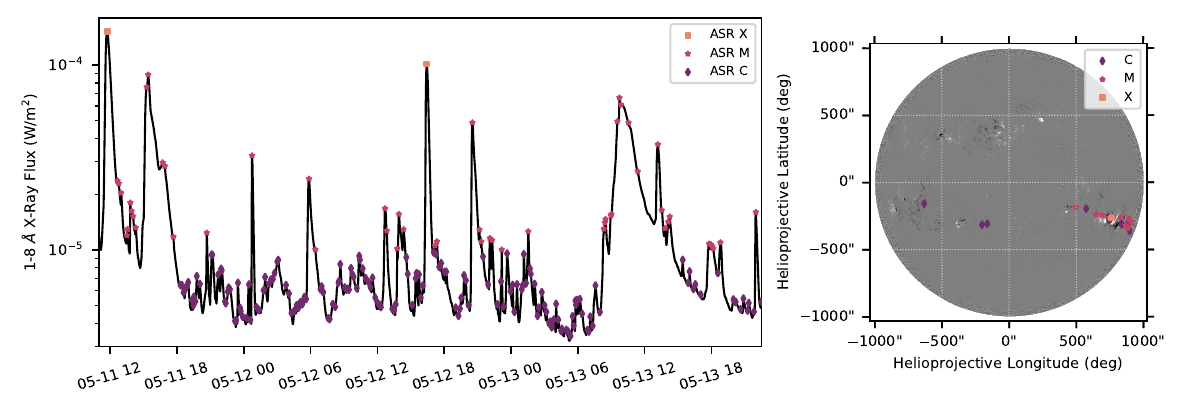}
    \caption{In the left panel, we show the measured GOES X-ray signal from May 11, 2024 to May 14, 2024. The detected flares are marked respectively in violet, pink, and orange diamonds for C, M and X class events. The right panel shows the positions of the most energetic flares in the solar disk over a magnetogram captured by the Helioseismic and Magnetic Imager onboard SDO on May 14, 2024. As expected, the majority of flares originated in a well-defined active region, while weaker flares are scattered across the solar disk.}
    \label{fig:3}    
\end{figure*}

In panel \textit{a)} of Fig.\ref{fig:1} we show the detection algorithm process for a particularly challenging case study. We selected the GOES X-ray signal from 9 March 2015, during the peak of solar activity of the 24th cycle, between 09:00 and 11:00 UTC, due to its complex features, which our algorithm successfully handled. The continuous black line is the 1-minute average of the soft X-ray flux in the long channel, and the star points represent the flare peaks. In particular, the pink stars highlight events that are identified both in the NOAA and ASR catalogues, whereas the orange ones are unique to our catalogue. The shaded regions mark the time of the identified flare events, and the dashed vertical line is the starting point. In the violet dashed box, we show that the algorithm successfully links two homologous events, as the second one starts before the first can return to pre-flare flux levels. The zoomed-inset provides an example of three consecutive extrema, which were identified as provisional maxima and minima. As explained previously, depending on the order of the sequence, this case may lead to a shift up to 2 minutes of the flare peak. In this case, the flare peak is attributed to the second maximum of the sequence.\\
The \textit{b)} and \textit{c)} panels of Fig. \ref{fig:1} present two statistical properties of our catalogue.
In panel \textit{b)}, we show the distribution of solar flares across the Sun's surface throughout the period covered by our catalogue. Similarly to sunspot behaviour, flares initially appear at mid-latitudes and migrate towards the equator as the solar cycle progresses. This pattern resembles the well-known butterfly diagram for sunspots \citep[e.g.][]{Cloutier2024}, but our analysis includes the positions of thousands of flares. The occurrence rate and the spatial distribution of flare events on the disk are correlated to solar activity, with higher latitudes seeing more flares when solar activity is higher. The solar activity minimum of the 24th solar cycle between 2018-2020 is also evident, as almost no intense flare events are observed.\\
We display the flare spatial distribution across the entire solar disk in panel \textit{c)}. Flares that belong to the M, C, and X standard classes are indicated by orange squares, pink crosses, and violet dots, respectively. Few stray values, which differ from the majority of the dataset constrained to mid-latitudes, are the result of the oversaturated AIA images. 

The probability density distribution of the flux integral and the peak flux for the ASR and NOAA catalogues are shown in Fig. \ref{fig:2}. The shapes of the distributions associated with the two datasets clearly differ at low energies. The flux integral distribution of the NOAA catalogue has a first peak around $10^{-8}$ W/m$^2$, followed by a minimum at $5\cdot10^{-7}$ W/m$^2$ and another maximum at $\sim$ $10^{-6}$ W/m$^2$. As XRS detection is effective for fluxes greater than $10^{-8}$ W / m$^2$, it could be argued that the first peak is driven purely by noise. Instead, the ASR probability distribution increases exponentially from lower energies to a peak around $\sim$ $10^{-6}$ W/m$^2$. This more "consistent" behaviour can be attributed to the effectiveness of the algorithm in identifying true flare events while discarding noise-driven fluctuations. At higher energies, both distributions are characterised by a similar exponential decay, covering more than two decades in energy. The dashed lines in Fig. \ref{fig:2} show the range in which we performed a linear fit of the distributions in the log-log space. The slopes associated with the two distributions are -2.1$\pm$0.1 and -1.9$\pm$0.1 for the ASR and NOAA catalogues, respectively. Our results are consistent with not only NOAA, but also with previous observations \citep[ASC, PLU, ][]{2002HvaOB..26....7V, 2002A&A...382.1070V, 2006ApJ...650L.143Y} and the scaling index of the fractal diffusive self-organised criticality (FD-SOC) model described in ASC and \cite{2011soca.book.....A}.

Most of the points discussed for the flux integral distribution can also be applied to the peak flux distribution shown in panel b) of Fig. \ref{fig:2}. Once again, the dashed lines are the linear fit of the distributions. The associated slopes are -2.3$\pm$0.1 and -2.1$\pm$0.1, which are consistent.

\subsection{May 2024 case study}

In this section, we present an in-depth focus on the most intense space weather event of the last two decades that occurred on 11-14 May 2024. The solar active regions classified as AR13664 and AR13668 by the NOAA, which eventually merged into a single one, produced a series of geo-effective space weather events comprising flares and fast CMEs. In the leftmost panel of Fig. \ref{fig:3}, we show the GOES X-Ray flux from May 11$^{th}$, 2024 to May 14$^{th}$, 2024, marking the peaks of the detected flare in violet, pink, and orange, respectively, for C, M, and X classes. According to the absolute class, i.e. the standard NOAA definition for flare class, our algorithm detected a total of 171 C-class flares, 53 M-class flares, and 2 X-class flares. In the right panel of Fig. \ref{fig:3} we show the position of the detected events on the solar disk. As we can see, the majority of flares originated from active regions 13664 and 13668 with a handful of less intense events coming from the other side of the solar disk. 

\section{Conclusion}

In this work, we presented the ASR catalogue, a novel flare catalogue based on GOES soft X-ray measurements that covers more than 20 years of observations. Based on an already established detection algorithm, we implemented significant changes to the identification algorithm and offered additional properties related to the flare, such as the point of origin of the most energetic events. By exploiting the user-friendliness of Python, we offer a robust and accessible alternative to already existing catalogues. The ASR catalogue offers reliable support for training machine learning models for space weather forecasting and solar physics studies. Finally, it is a live service that provides a daily bulletin of flares in the past 24 hours. 

\section*{Data Availability}
The ASR catalogue and its code can be freely accessed at the following link: \url{https://github.com/helio-unitov/ASR_cat}

\vspace{0.5cm}

\begin{acknowledgments}
    
We acknowledge the use of data from GOES satellites, concerning solar signal in the soft-X range (\url{https://www.ngdc.noaa.gov/stp/satellite/goes/dataaccess.html}) and the use of NOAA SWPC flare lists (\url{https://www.ngdc.noaa.gov/stp/solar/solarflares.html}).
MB acknowledges that this article was produced while attending the PhD program in Space Science and Technology at the University of Trento, Cycle XXXIX, with the support of a scholarship financed by the Ministerial Decree no. 118 of 2nd March 2023, based on the NRRP - funded by the European Union - NextGenerationEU - Mission 4 ``Education and Research", Component 1 ``Enhancement of the offer of educational services: from nurseries to universities” - Investment 4.1 “Extension of the number of research doctorates and innovative doctorates for public administration and cultural heritage” - CUP E66E23000110001.
S.M. acknowledges the Ph.D. course in Astronomy, Astrophysics and Space Science of the University of Rome ``Sapienza”, University of Rome ``Tor Vergata” and Istituto Nazionale di Geofisica e Vulcanologia, Italy.
\end{acknowledgments}

\vspace{5mm}


\bibliography{asr}{}
\bibliographystyle{aasjournal}

\begin{appendix}

    \section{Background Characterisation}\label{sec:app}

\begin{figure*}[ht!]
        \centering
        \includegraphics[width=\linewidth]{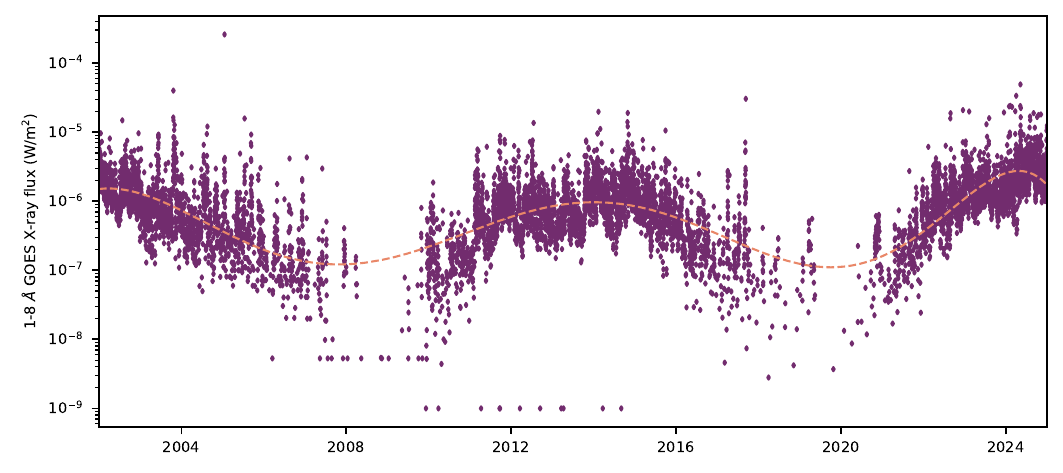}
        \caption{Observed background flux over the entire duration of the dataset. The orange dashed line shows a seventh-degree polynomial fit. For clarity, we included only one event out of every 30 to avoid overcrowding with overlapping flares.}
        \label{fig:app}
    \end{figure*}

    In Fig. \ref{fig:app} we show the observed flux background over the full temporal length of our dataset. The clear reproduction of the solar activity cycle emphasises the importance of accounting for background variations when estimating the event integral, ensuring an unbiased measure of flare energetics. The concentration of points at lower energies represents the full scale of the instrument.

    \section{Data product description}
    The head of the ASR catalogue is organized in 13 columns. A full explanation for each tag is given in the paper, here we provide a table with a brief description of each column. 

    \begin{table}[h!]
    \centering
    \begin{tabular}{|c|l|}
        \hline
        \textbf{Column} & \textbf{Description} \\  
        \hline
        \texttt{flare\_id} & Identificative number of a flare; not unique as consecutive flares share the same ID. \\  
        \texttt{tstart} & Flare starting time in ISO8601 datetime format. \\  
        \texttt{tpeak} & Flare peak time in ISO8601 datetime format. \\  
        \texttt{tend} & Flare end time in ISO8601 datetime format. \\
        \texttt{BG\_flux} & Background flux computed over the two-minute interval before the event. \\  
        \texttt{flux\_integral} & Flux integral of the flare event from \texttt{tstart} to \texttt{tend}. \\  
        \texttt{ratio} & Ratio between background and peak fluxes. \\  
        \texttt{flux\_int\_corrected} & Flux integral from \texttt{tstart} to \texttt{tend}, accounting for background flux. \\  
        \texttt{delta\_flux} & Difference between peak flux and background flux. \\  
        \texttt{fclass\_simple} & Flare class estimated based on \texttt{delta\_flux}, accounting for background flux. \\  
        \texttt{fclass\_full} & Full class of the flare estimated as in \texttt{fclass\_simple}. \\  
        \texttt{abs\_class\_simple} & Flare class estimated according to NOAA. \\  
        \texttt{abs\_class\_full} & Full class of the flare estimated as in \texttt{abs\_class\_simple}. \\  
        \hline
    \end{tabular}
    \caption{Description of the columns included in the flare catalogue.}
    \label{tab:flare_catalogue}
\end{table}

    Furthermore, the three catalogues that also report the position of the flare events (one for each class) have six additional columns that accommodate the flares coordinates in different reference frames.

        \begin{table}[h!]
    \centering
    \begin{tabular}{|c|l|}
        \hline
        \textbf{Column} & \textbf{Description} \\  
        \hline
        \texttt{flare\_x} & Position of the flare in pixel coordinates \\  
        \texttt{flare\_y} & Position of the flare in pixel coordinates \\  
        \texttt{Lon [Helioprojective]} & Longitude of the flare in the helioprojective reference frame \\  
        \texttt{Lat [Helioprojective]} & Longitude of the flare in the helioprojective reference frame \\  
        \texttt{Lon [Heliographic]} & Longitude of the flare in the heliographic Stonyhurst reference frame \\  
        \texttt{Lat [Heliographic]} & Longitude of the flare in the heliographic Stonyhurst reference frame \\  
        \hline
    \end{tabular}
    \caption{Description of the additional columns included in the flare catalogues with the positions.}
    \label{tab:flare_catalogue}
\end{table}
      
\end{appendix}

\end{document}